\documentclass[twocolumn,showpacs,preprintnumbers,amsmath,amssymb,prl,aps,superscriptaddress]{revtex4-1}
\usepackage{graphicx}
\usepackage{amsmath}
\usepackage{verbatim}

\newcommand{\Ma}{M_0}
\newcommand{\Mb}{M_1}
\newcommand{\Mc}{M_2}
\newcommand{\Fbg}{\mathrm{Fb}_0}
\newcommand{\Fbe}{\mathrm{Fb}_1}
\newcommand{\Pg}{P_{\ket{0}}}
\newcommand{\Pe}{P_{\ket{1}}}
\newcommand{\Pee}{P_{\ket{2}}}
\newcommand{\Gzo}{\Gamma_{01}}
\newcommand{\Goz}{\Gamma_{10}}
\newcommand{\Got}{\Gamma_{12}}
\newcommand{\Gto}{\Gamma_{21}}

\newcommand{\tinit}{\tau_{\mathrm{init}}}
\newcommand{\Perr}{P_{\mathrm{err}}}

\newcommand{\Vps}{V_{\mathrm{ps}}}
\newcommand{\tlat}{\tau}
\newcommand{\us}{\mu\mathrm{s}}
\newcommand{\mK}{\mathrm{mK}}
\newcommand{\ket}[1]{\left\lvert #1 \right\rangle}
\newcommand{\wzo}{\omega_{01}}
\newcommand{\GHz}{\mathrm{GHz}}
\newcommand{\wrr}{\omega_{\mathrm{r}}}
\newcommand{\wrf}{\omega_{\mathrm{RF}}}
\newcommand{\kHz}{\mathrm{kHz}}
\newcommand{\MHz}{\mathrm{MHz}}
\newcommand{\ns}{\mathrm{ns}}
\newcommand{\Vthr}{V_{\mathrm{th}}}
\newcommand{\zotrans}{0\leftrightarrow1}

\begin{document}
\title{Feedback control of a solid-state qubit using high-fidelity projective measurement}
\author{D.~Rist\`e}
\author{C.~C.~Bultink}
\affiliation{Kavli Institute of Nanoscience, Delft University of Technology, P.O. Box 5046,
2600 GA Delft, The Netherlands}
\author {K.~W.~Lehnert}
\affiliation{JILA, National Institute of Standards and Technology and the University of Colorado and Department of Physics, University of Colorado, Boulder, Colorado 80309, USA}
\author{L.~DiCarlo}
\affiliation{Kavli Institute of Nanoscience, Delft University of Technology, P.O. Box 5046,
2600 GA Delft, The Netherlands}
\date{\today}

\begin{abstract}
We demonstrate feedback control of a superconducting transmon qubit using discrete,  projective measurement and conditional coherent driving. Feedback realizes a fast and deterministic qubit reset to a target state with  $2.4\%$ error averaged over input superposition states, and  cooling of the transmon from $16\%$ spurious excitation to $3\%$. This closed-loop qubit control  is necessary for measurement-based protocols such as quantum error correction and teleportation.
\end{abstract} 

\pacs{03.67.Lx, 42.50.Dv, 42.50.Pq, 85.25.-j}
\maketitle

Many protocols in quantum information processing (QIP) require closing a feedback  loop where coherent control of qubits is conditioned on projective measurements in real time~\cite{Wiseman09}. Important examples include quantum error correction and teleportation~\cite{Nielsen00}, so far achieved in ion-trap~\cite {Chiaverini04,Riebe04,*Barrett04} and photonic systems~\cite{Pittman05,Furusawa98}. 
During the last decade, the steady development of qubit readout and universal gates needed in a quantum processor~\cite{DiVincenzo00} has made superconducting circuits~\cite{Clarke08} a leading solid-state QIP platform. However, the simple quantum algorithms~\cite{DiCarlo09,*Yamamoto10,*Dewes12, *Lucero12} and teleportation-like protocol~\cite{Baur12} so far demonstrated fall in the category of open-loop control. Measurement is performed as the final step, following a programmed sequence of applied gates. A comparable realization of closed-loop control has been precluded by stringent requirements on high measurement fidelity and short loop delay (latency).  
Until recently, the available qubit coherence times  bottlenecked both achievable fidelity and required speed. 

For feedback control of superconducting qubits, the development of  circuit quantum electrodynamics~\cite{Blais04,Wallraff04} with 3D cavities (3D cQED)~\cite{Paik11} constitutes a watershed. The new order of magnitude in qubit coherence times $(>10~\us$) allows boosting projective-readout fidelity up to $98\%$~\cite{Riste12,Johnson12} and realizing feedback with off-the-shelf electronics. Very recently, feedback based on continuous weak measurement has sustained Rabi oscillations of a transmon qubit indefinitely~\cite{Vijay12}. Previously, this type of feedback had only been used to generate and stabilize quantum states of  photons~\cite{Gillett10,*Sayrin11}, ions~\cite{Bushev06}, and atoms~\cite{Koch10,*Brakhane12}.

\begin{figure}[b]
\includegraphics[width=0.8\columnwidth]{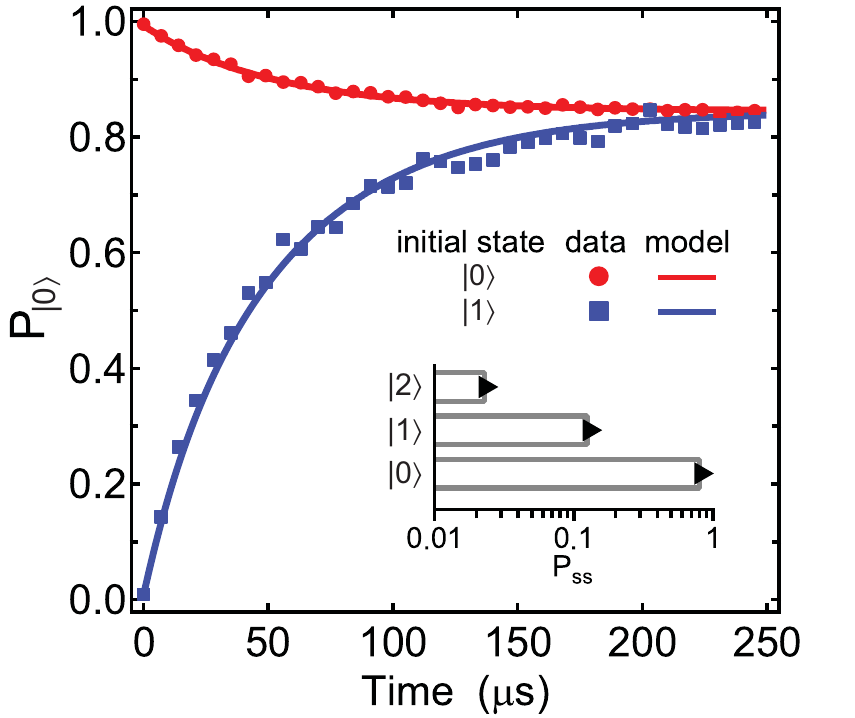}
\caption{(color online). Transmon equilibration to steady state. Time evolution of the ground-state population $\Pg$ starting from states $\ket{0}$ and $\ket{1}$. Solid curves are the best fit (including data in Fig.~S4) to Eq.~1, giving the inverse transition rates $\Gzo^{-1 }= 50\pm2~\us, \Got^{-1} = 20\pm2~\us, \Goz^{-1} = 324\pm32~\us, \Gto^{-1} = 111\pm25~\us$. From the steady-state solution, we extract residual excitations $P_{\ket{1},\mathrm{ss}}=13.1\pm0.8\%,P_{\ket{2},\mathrm{ss}}=2.4\pm0.4\%$. Inset: steady-state population distribution (bars). Markers correspond to a Boltzmann distribution with best-fit  temperature $127~\mK$, significantly higher than the dilution refrigerator base temperature ($15~\mK).$}
\end{figure}

In this Letter, we demonstrate  feedback control of a superconducting transmon qubit based on discrete, projective measurement. This dual type of feedback is the kind necessary for measurement-based QIP. As a first application, we demonstrate a feedback-based reset that is deterministic and fast compared to passive initialization. This feedback cools the transmon from a spurious steady-state excitation of $16\%$ to $3\%$ and resets qubit states with $2.4\%$ error averaged over the Bloch sphere. These absolute errors are dominated by latency, in quantitative agreement with a model including transmon equilibration and readout errors.  

The experiment employs a transmon qubit inside an aluminum 3D cavity~\cite{Paik11}.  The qubit ($\wzo/2\pi = 4.889~\GHz$ transition frequency) couples  to the cavity fundamental mode ($\wrr/2\pi=6.546~\GHz$, coupling-limited linewidth $\kappa/2\pi=550~\kHz$) with strength $g/2\pi=68~\MHz$.  The high-fidelity, projective qubit readout forming the input to the feedback loop uses homodyne detection of the qubit-state dependent cavity transmission (dispersive shift $2\chi/2\pi =-1.9~\MHz$~\cite{Wallraff04}). A $400~\ns$ measurement pulse at $\wrf = \wrr - \chi$ is applied to the cavity and the transmitted signal is then amplified by a Josephson parametric amplifier~\cite{Castellanos-Beltran08, Vijay09} to enhance sensitivity~\cite{SOM}. The feedback controller is an ADwin Gold processor that samples the transmitted homodyne signal, performs 1-bit digitization to interpret the projected qubit state, and conditionally triggers a $\pi$ pulse resonant with the transmon $0\leftrightarrow1$ transition. The $2.64~\us$ delay between start of the measurement and end of the $\pi$ pulse, set by processing time in the ADwin, is  short compared to the qubit relaxation time $T_1$ (see {below})~\cite{shortT2b}. 

Our first application of feedback is qubit initialization, also known as reset~\cite{DiVincenzo00}. To be useful in QIP, reset must be deterministic (as opposed to heralded or postselected~\cite{Riste12, Johnson12}) and fast compared to qubit coherence times.
Obviously, the passive method, i.e., waiting several $T_1$ times, does not meet the speed requirement. Moreover, it can suffer from residual steady-state qubit excitations~\cite{Corcoles11,Johnson12,Riste12,Vijay12}, whose source remains an active area of research. The drawbacks of passive initialization are evident for our qubit, whose ground-state population $\Pg$ evolves from states  $\ket{0}$ and $\ket{1}$ as shown in Fig.~1. The transmon dynamics are captured by a master equation model for a three-level system~\cite{Bianchetti09,SOM}:
\begin{equation}
\
\left(\begin{array}{c}
\dot{P}_\mathrm{\ket{0}}\\
\dot{P}_\mathrm{\ket{1}}\\
\dot{P}_\mathrm{\ket{2}}
\end{array}\right)=
\left( \begin{array}{ccc}
-\Goz & \Gzo & \phantom{-}0 \\
\phantom{-}\Goz & -\Gzo-\Gto & \phantom{-}\Got \\
\phantom{-}0 & \Gto & -\Got
\end{array} \right)
\left(\begin{array}{c}
P_\mathrm{\ket{0}}\\
P_\mathrm{\ket{1}}\\
P_\mathrm{\ket{2}}   
\end{array}\right).
\end{equation} 
The best fit to the data gives the qubit relaxation time $T_1=1/\Gzo=50\pm2~\us$ and $15.5\%$ residual total excitation. 
Previous approaches to accelerate qubit equilibration include coupling to dissipative resonators~\cite{Reed10b} or two-level systems~\cite{Mariantoni11}. However, these are also susceptible to spurious excitation, potentially inhibiting complete qubit relaxation.

\begin{figure}
\includegraphics[width=0.89\columnwidth]{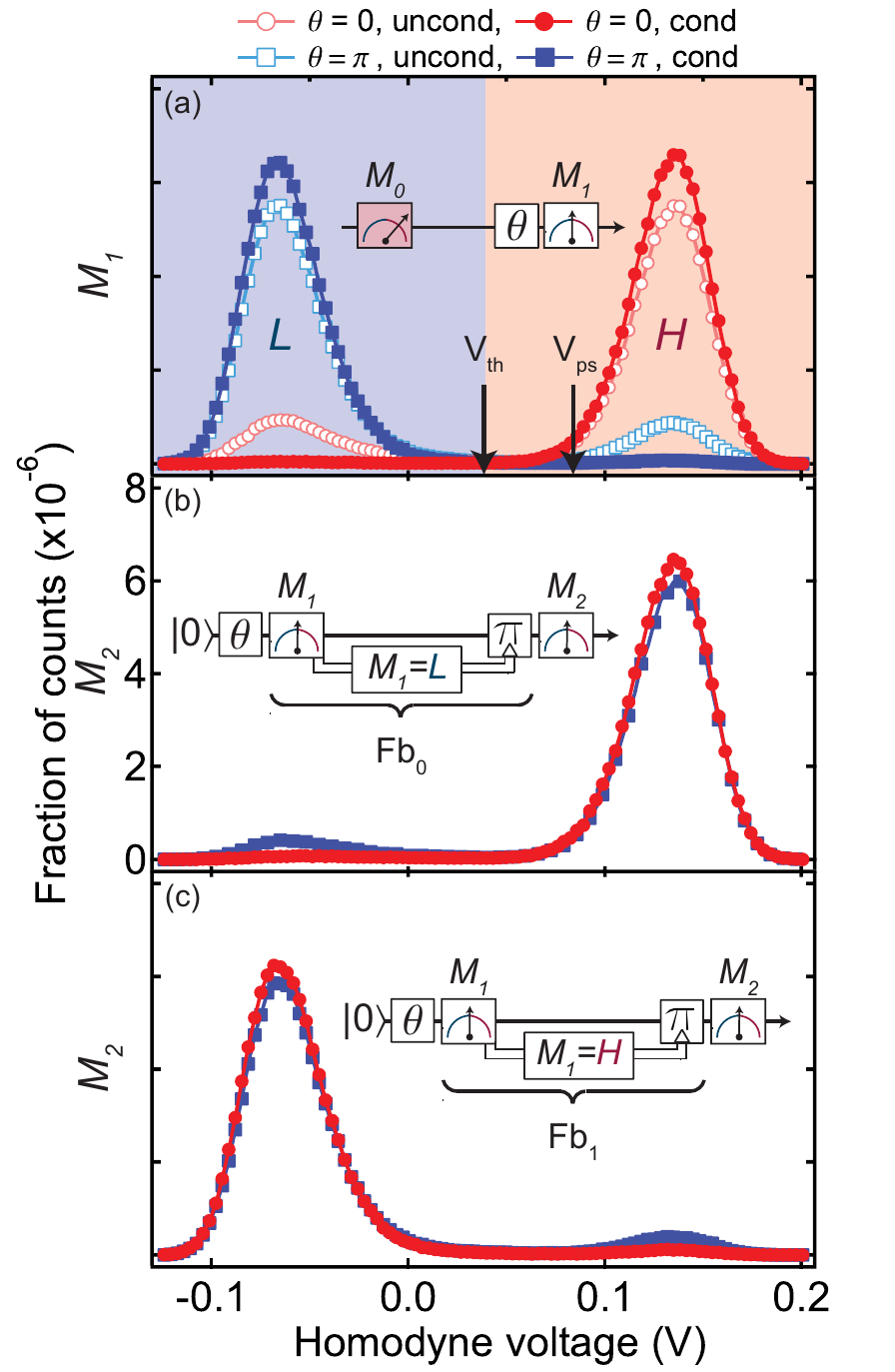}
\caption{(color online). Reset by measurement and feedback. (a) Histograms of $300\,000$ shots of $\Mb$, with (squares) and without (circles) inverting the qubit population with a $\pi$ pulse. $H$ and $L$ denote the two possible outcomes of $\Mb$, digitized with the threshold $V_\mathrm{th}$, maximizing the contrast.  Full (empty) dots indicate (no) postselection on $\Ma>\Vps$, chosen to
prepare $\ket{0}$ with $0.7\%$ error while keeping $81\%$ of the data. This technique purifies the steady-state mixture (see Fig.~1), preparing the input states for the feedback sequences in (b) and (c).
(b) Histograms of $\Mc$  after applying the feedback protocol $\Fbg$, which triggers a $\pi$ pulse when $\Mb=L$. Using this feedback,  $\sim99\%$ (92\%) of counts fall in the $H$ area for $\theta=0~(\pi)$, respectively. (c) Feedback with opposite logic $\Fbe$ preparing the excited state. In this case, $\sim 98\%$ $(94\%)$ of counts fall in the $L$ area for $\theta=0~(\pi)$.}
\end{figure}

Feedback-based reset circumvents the equilibration problem by not relying on coupling to a dissipative medium. Rather, it works by projecting the qubit with a measurement ($\Mb$, performed by the controller) and conditionally applying a $\pi$ pulse to drive the qubit  to a targeted basis state (Fig.~2). 
A final measurement ($\Mc$) determines the qubit state immediately afterwards, by averaging the second half ($200~\ns)$ of a transmitted measurement pulse sampled by a data acquisition card~\cite{SOM}. 
In both measurements, the result is digitized into levels $H$ or $L$, associated with $\ket{0}$ and $\ket{1}$, respectively. The digitization threshold voltage $\Vthr$ maximizes the readout fidelity at $99\%$. 
The $\pi$ pulse is conditioned on $\Mb=L$ to target $\ket{0}$ (scheme $\Fbg$) or on $\Mb=H$ to target $\ket{1} (\Fbe)$. To benchmark the reset protocol, we quantify its action on the purest states we can prepare. This step is accomplished with a preliminary  measurement $\Ma$ (initializing the qubit in $\ket{0}$ by postselection~\cite{Riste12,Johnson12}), followed by a calibrated pulse resonant on the transmon $\zotrans$ transition to prepare $\ket{\theta}=\cos(\theta/2)\ket{0}+\sin({\theta/2})\ket{1}$. We first consider the action of reset on the basis states ($\theta=0,$ $\pi$). The overlap of the $\Mc$ histograms with the targeted region ($H$ for $\Fbg$ and $L$ for $\Fbe$) averages at $96\%$, indicating the success of reset. Imperfections are more evident for $\theta=\pi$ and mainly due to equilibration of the transmon during the feedback loop. We address this point quantitatively below. 

\begin{figure}
\includegraphics[width=\columnwidth]{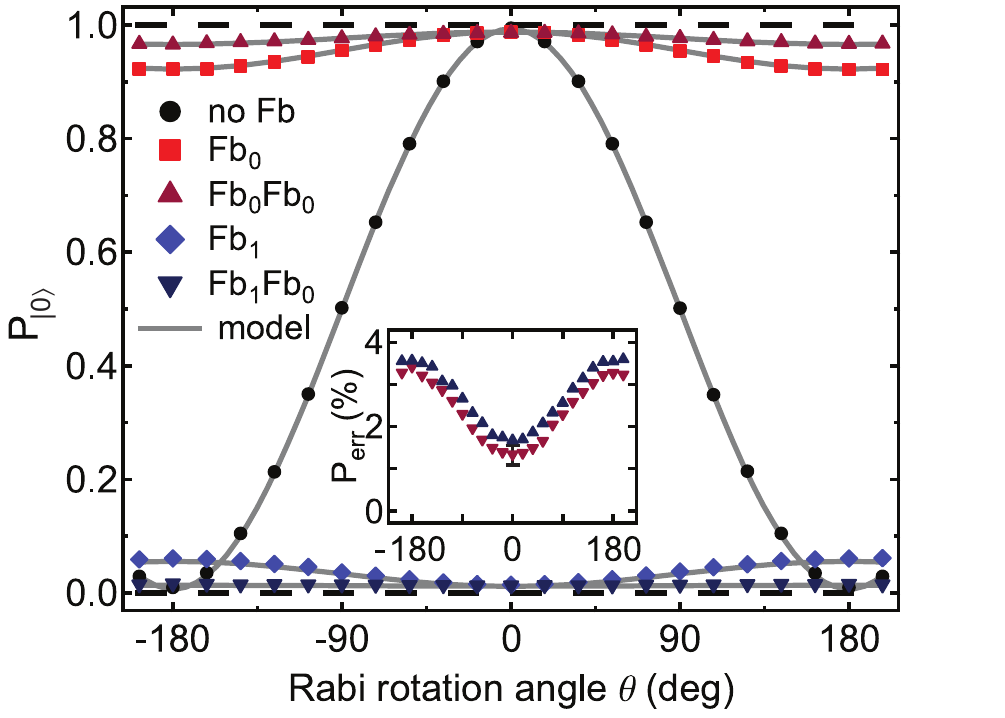}
\caption{(color online). Deterministic reset from any qubit state.
Ground-state population $\Pg$ as a function of the initial state $\ket{\theta}=\cos(\theta/2)\ket{0}+\sin({\theta/2})\ket{1}$, prepared by coherent rotation after initialization in $\ket{0}$, as in Fig.~2. The cases shown are: no feedback (circles), one instance of $\Fbg$ (squares) or $\Fbe$ (diamonds), and two instances of $\Fbg$ (upward triangles) or $\Fbg$ followed by $\Fbe$ (downward triangles). 
Inset: error probabilities for two rounds of feedback, defined as $1-P_{\ket{t}}$, where $\ket{t}\in\{0,1\}$ is the target state. The vertical axis is calibrated with the reference states (Fig.~2) and corrected for the partial excitation between $\Ma$ and $\Mb$. The systematic $\sim0.3\%$ difference between the two cases is attributed to error in the $\pi$ pulse preceding the measurement of $\Pe$ following $\Fbe$~\cite{SOM}.  Curves: model including readout errors and equilibration~\cite{SOM}.}
\end{figure}

An ideal reset function always prepares the same pure qubit state, regardless of its  input. To fully quantify the performance of our reset scheme, we measure its effect on different superposition states $\ket{\theta}$. 
After applying $\Fbg$ $(\Fbe)$, we extract $\Pg$ by averaging over $300\,000$ repetitions of the experiment (Fig.~3).
Without feedback, $\Pg$ is trivially a sinusoidal function of $\theta$, with contrast slightly reduced from unity due to a preparation infidelity of $0.7\%$.
With feedback, we observe a suppression of the Rabi oscillation, with $\Pg$ approaching the ideal value 1 (0) for $\Fbg$  $(\Fbe)$ for any input state. However, a dependence on $\theta$ remains, with $\Perr=1-\Pg$ for $\Fbg$ ($1-\Pe$ for $\Fbe$) ranging from $1.2\%$ $(1.4\%)$ for $\theta=0$ to $7.8\%$ ($8.4\%$) for $\theta = \pi$. The remaining errors have two sources: mismatch between measurement result $M$ and post-measurement state $\ket{i}$, occurring with probability $p^M_{ij}$, for initial state $\ket{j}$; and equilibration during the \ $\tlat=2.4~\us$ lapse between the end of $\Mb$ and the start of $\Mc$, set by processing time in the controller. 
Transitions to $\ket{2}$ during $\Mb$ (with probability  $p_{21}=p^H_{21}+p^L_{21}$), or during $\tlat$ ($\Gamma_{21}\tlat$), cause leakage out of the qubit subspace, where the feedback has no action. 
For perfect pulses, the overall errors (equal for $\Fbg$ and $\Fbe$) are to first order:
\begin{align*}
&\Perr^{\theta=0} = p^L_{00}+p^H_{10}+\Gamma_{10}\tlat, \\
&\Perr^{\theta=\pi}=  p^H_{11}+p^L_{01}+p_{21}+(\Gamma_{01}+\Gamma_{21})\tlat,
\end{align*}
and weighted combinations thereof for other $\theta$. 
Using the best-fit rates $\Gamma_{ij}$ (Fig.~1), errors due to equilibration sum to $0.7\%$ $(6.9\%)$ for $\theta=0$ $(\pi)$, while readout errors account for the remaining $0.4\%$ $(1.4\%)$. 

To improve reset fidelities, we concatenate two feedback cycles. The dominant error for $\theta=0$  remains unchanged, but for $\theta=\pi$ it decreases to $\Perr^{\theta=0}+p_{21}+\Gamma_{21}\tlat$, in agreement with the measured values of $1.3\%$ and $3.4\%$, respectively.  The consecutive application of $\Fbg$ and $\Fbe$ prepares $\ket{1}$ with similar fidelities. 

\begin{figure}
\includegraphics[width=\columnwidth]{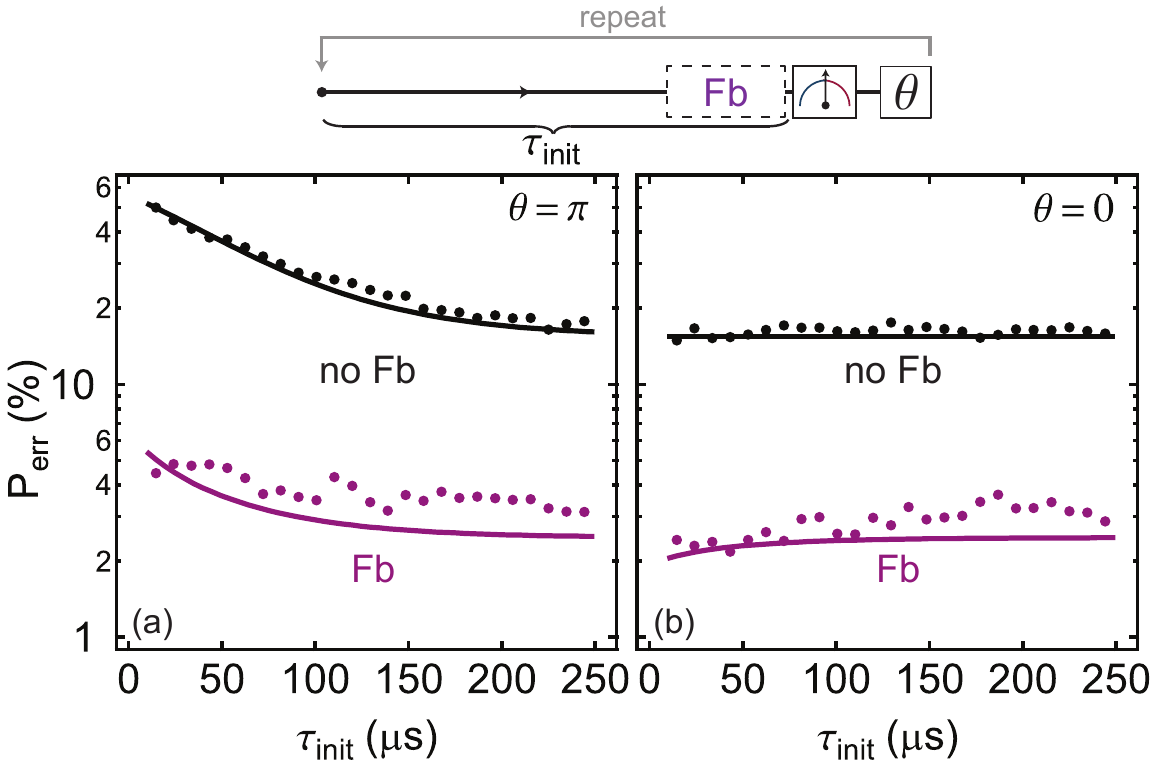}
\caption{(color online). Fast qubit reset. Initialization errors under looped execution of a simple experiment consisting of (a) a $\pi$ pulse and (b) no pulses, leaving the qubit ideally in $\ket{1}$ or $\ket{0}$, respectively. Subsequent runs are separated by an initialization time $\tinit$. Black: initialization by waiting (no feedback). Purple: initialization by feedback, with three rounds of $\Fbg$ and a $\pi$ pulse on the $1\leftrightarrow2$ transition~\cite{SOM}. Curves correspond to a master equation simulation assuming perfect pulses.}
\end{figure}

The key advantage of reset by feedback is the ability to ready a qubit for further computation fast compared to coherence times available in 3D cQED~\cite{Paik11,Rigetti12}. This will be important for ancilla qubits used for multiple rounds of error correction~\cite{Schindler11}.  
To test the gain in speed over passive initialization, we run a sequence multiple times, separated by a variable initialization time $\tinit$, during which we perform reset or not (Fig.~4). A measurement pulse follows the initialization period to quantify the initialization error $\Perr$. The sequence itself emulates an algorithm leaving the qubit in $\ket{1}$ (or $\ket{0}$), by simply applying (or not applying) a $\pi$ pulse. Without feedback, when the sequence is a $\pi$ pulse [Fig.~4(a)] the qubit remains partially excited and $\Perr$ approaches $50\%$ as $\tinit \to 0$. With no $\pi$ pulse [Fig.~4(b)], $\Perr$ simply equals the total steady-state excitation. Feedback during $\tinit$ suppresses the reset error from $\sim32\%$ (averaged between the two cases) to $3.5\%$ for the shortest $\tinit$ ($15~\us$). We achieve this improvement by combining three rounds of $\Fbg$ with a $\pi$ pulse on the  $1\leftrightarrow2$ transition before the final $\Fbg$, which partially counters leakage to the second excited state~\cite{SOM}. In the opposite limit $\tinit \gg T_1$, residual excitations are reduced from $16\%$ to $3\%$~\cite{nonthermal}, cooling the transmon. 

In conclusion, we have demonstrated feedback control of a transmon qubit using high-fidelity projective measurement and conditional operation. We have applied this feedback to deterministically reset the qubit, starting from any superposition, with an average error of $2.4\%$. We have also used feedback-based initialization to loop a test sequence with more than $15$-fold increase in repetition rate relative to the passive method. Reset fidelity is currently limited by transitions to higher energy levels, caused by the $127~\mK$ transmon temperature. We estimate that by reducing this temperature below $50~\mK$, the suppression of $\Gamma_{10}$ and $\Gamma_{21}$ will decrease the reset error past the $\sim1\%$ fault-tolerant threshold in surface coding~\cite{Wang11}. This reduction may be achieved by a combination of infrared radiation shielding~\cite{Corcoles11}, use of a copper cavity~\cite{Cuexc}, and improved qubit thermal anchoring.  Moreover, decreasing the latency in the feedback loop, for example by using field-programmable gate arrays, will further reduce errors due to equilibration between measurement and action. While this demonstration employs feedback on a single qubit, the scheme can be extended to conditionally drive another qubit within its coherence time, realizing the feedforward~\cite{Wiseman09} needed for teleportation and measurement-based error correction.  Future experiments will also target the generation of entanglement by combining the demonstrated feedback scheme with cavity-based parity measurements~\cite{Lalumiere10, Tornberg10}.

\begin{acknowledgments}
We thank  V.~Ranjan, J.~G.~van Leeuwen, and H.-S.~Ku for experimental assistance, and M.~Tiggelman,  R.~N.~Schouten, and A.~Wallraff for discussions. We acknowledge funding from the Dutch Organization for Fundamental Research on Matter (FOM), the Netherlands Organization for Scientific Research (NWO, VIDI scheme), the EU FP7 project SOLID, and the DARPA QuEST program.
\end{acknowledgments}
{\it Note added.}---A parallel manuscript from \'Ecole Normale Sup\'erieure, Paris~\cite{CampagneIbarcq12} reports feedback control of a transmon qubit using a Josephson parametric converter for high-fidelity projective readout.


\begin{thebibliography}{10}
\bibitem{shortT2b}
The Ramsey dephasing time $T_2^*=2~\us$ is most likely due to photon shot
  noise~\cite{Sears12} and sensitivity to charge noise due to the low ratio
  $E_J/E_C~\sim 27$~\cite{Schreier08}. Dephasing is however not relevant for
  the reset application here demonstrated.
\bibitem{nonthermal}
We note that $\Pe\approx\Pee=1.6\%$ is a non-thermal distribution.
\bibitem{Cuexc}
We have recently measured a residual excitation of $\sim2\%$ for a qubit at the
  same frequency inside a Cu cavity of the same geometry, consistent with
  Ref.~\onlinecite{Rigetti12}.
\end{thebibliography}
\end{document}


\title{Supplement to ``Feedback control of a solid-state qubit using high-fidelity projective measurement"}
\author{D.~Rist\`e}
\author{C.~C.~Bultink}
\affiliation{Kavli Institute of Nanoscience, Delft University of Technology, P.O. Box 5046,
2600 GA Delft, The Netherlands}
\author {K.~W.~Lehnert}
\affiliation{JILA, National Institute of Standards and Technology and the University of Colorado and Department of Physics, University of Colorado, Boulder, Colorado 80309, USA}
\author{L.~DiCarlo}
\affiliation{Kavli Institute of Nanoscience, Delft University of Technology, P.O. Box 5046,
2600 GA Delft, The Netherlands}
\date{\today}
\maketitle

\section{Readout and feedback configurations}
The readout frequency is set to $\wrf=\wrr - \chi$ to maximize the phase difference in cavity transmission between qubit in $\ket{0}$ and $\ket{1}$. The JPA is tuned to $\wrf$ by the combination of external flux biasing and a continuous-wave pump tone that bends the JPA resonance, as described in Ref.~\onlinecite{Riste12}. In contrast to this previous work, here we use a single generator for pump, measurement and local oscillator in order to maximize phase stability. A phase shifter on the RF arm is used to maximize the phase-sensitive amplification [Fig.~S3(a)]. Another phase shifter at the output side maximizes the in-phase signal after IQ demodulation. An additional circulator between cavity and JPA is used to further suppress leakage of the pump into the cavity (Fig.~S1).
\begin{figure*}
\includegraphics[width=1.44\columnwidth]{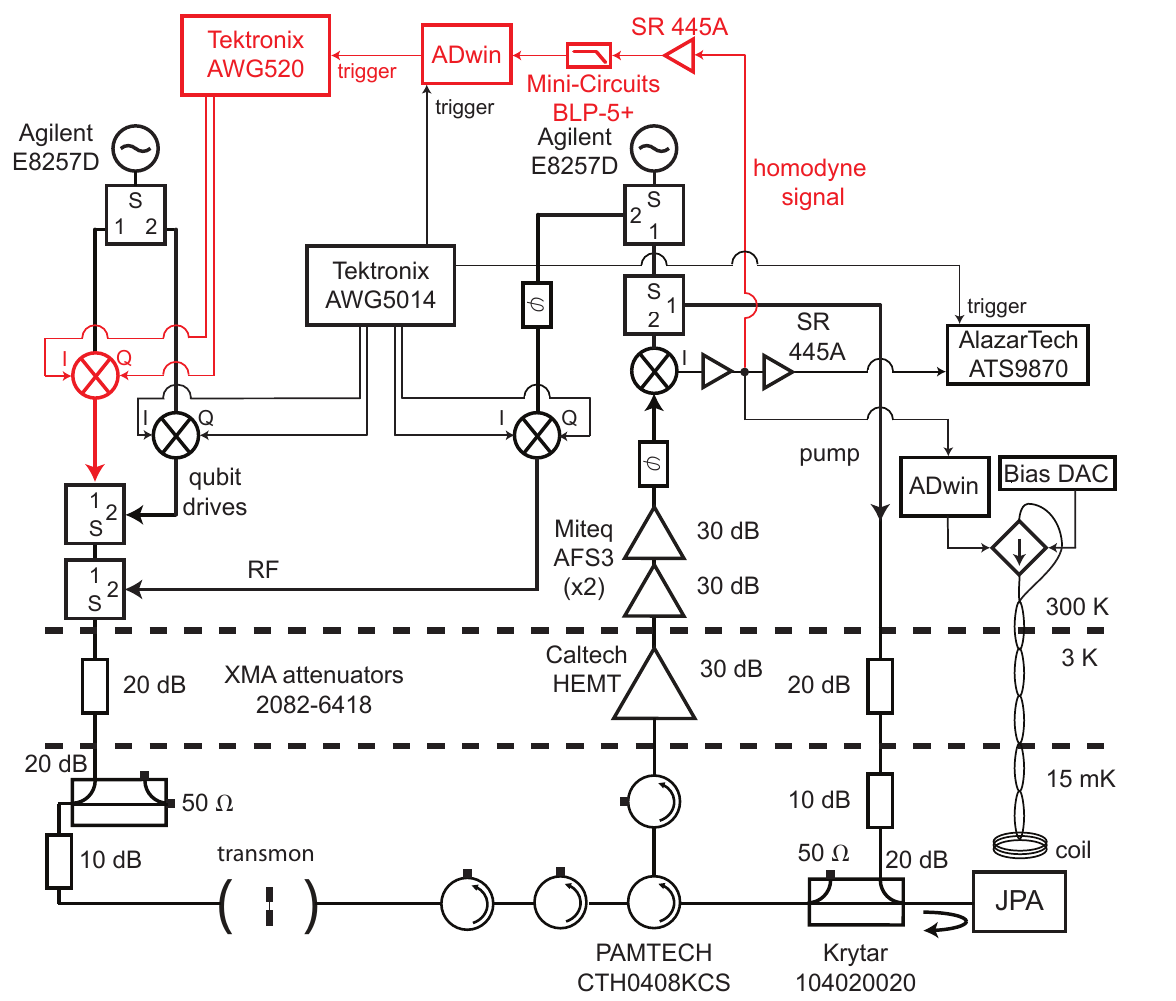}
\caption{Experimental setup. The key components of the feedback loop are highlighted in red. A Tektronix AWG5014 generates modulation envelopes for measurement and deterministic qubit pulses. The readout signal is amplified with a JPA at base temperature, a HEMT (Caltech CITCRYO 1-12) at $3~\K$, and two GaAs FET amplifiers (Miteq AFS3) at room temperature prior to demodulation to baseband ($0~\Hz$ intermediate frequency). The demodulated signal is amplified (Stanford Research Systems SR445A) and then split. One arm is sent through a low-pass filter (Mini-Circuits BLP-5+, DC-$5~\MHz$ passband) to an analog input of an ADwin Gold processor, and the other to the data acquisition card (AlazarTech). The ADwin compares the input voltage to the optimum threshold for discriminating between $\ket{0}$ and $\ket{1}$. Depending on the feedback logic programmed into it, the ADwin sends a trigger to the Tektronix AWG520 when the input is above/below the threshold. The triggered AWG520 produces the modulation envelope for a $\pi$ pulse sent via the input line, closing the feedback loop.}
\end{figure*}

The ADwin samples the readout signal once, at a variable delay following a trigger from an arbitrary waveform generator (Tektronix AWG5014). This delay is adjusted in $100~\ns$ steps to maximize readout contrast between states prepared with and without a $\pi$ pulse on the $\zotrans$ transition. For multiple feedback instances, the timing for each measurement is separately optimized. Once these timings are set, a routine (repeated every $\sim5$~min), determines the optimum threshold for digitizing the readout signal. This voltage is then used to assign $H$ or $L$ to the measurement in any feedback cycle. For $\Fbg~(\Fbe)$, the ADwin triggers another arbitrary waveform generator  (Tektronix AWG520) to produce a $\pi$ pulse when the outcome is $L~(H)$. Pulse timings and signal delays in the feedback cycle are illustrated in Fig.~S2.

\begin{figure*}
\includegraphics[width=2\columnwidth]{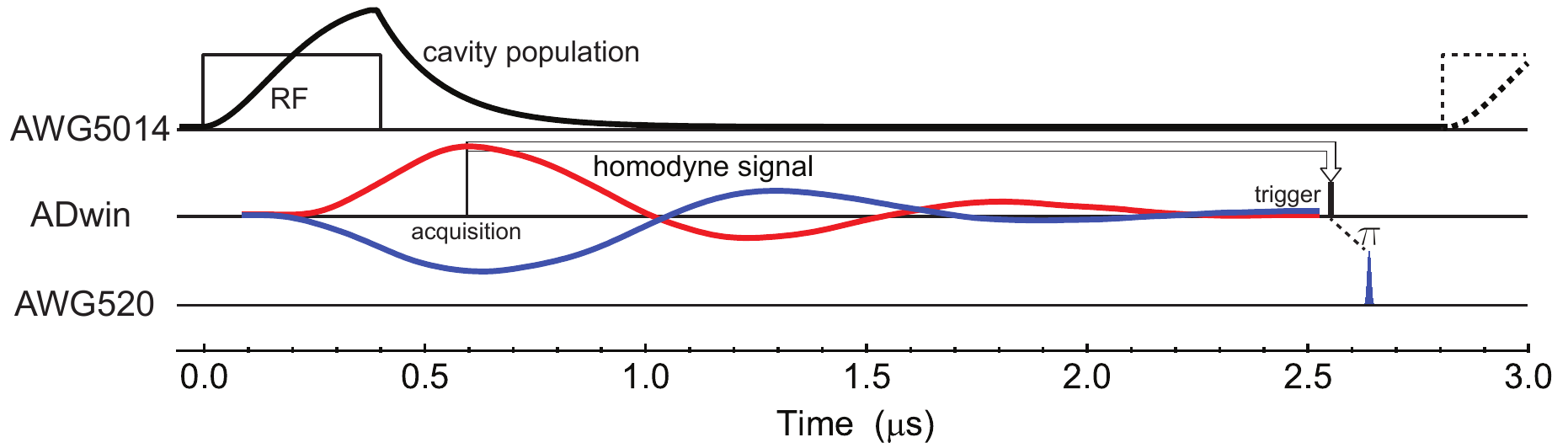}
\caption{Feedback loop latency. The AWG5014 modulates the RF tone to generate a $400~\ns$ measurement pulse reaching the cavity at $t=0$. The black curve is a sketch of the photon population inside the cavity, which reaches $\sim45$ photons at $t=0.4~\us$. We estimate a critical photon number $n_\mathrm{crit}\approx150$ and inverse measurement rate $\Gamma_m^{-1}=26~\ns$ at steady state~\cite{Gambetta06}. The ADwin, triggered by the AWG5014, measures one channel of the output homodyne signal (red: qubit in $\ket{0}$, blue: $\ket{1}$), delayed by $\sim200~\ns$ due to the low pass filter at its input side. After comparison of the measured voltage at $t=0.6~\us$ to the reference threshold, the AWG520 is conditionally triggered at $t=2.54~\us$, resulting in a $\pi$ pulse (Gaussian, $\sigma=6~\ns$, total length $24~\ns$), reaching the cavity at $2.62~\us$. The ADwin acquisition (and the triggered $\pi$ pulse) can be shifted by $\pm0.1~\us$ to maximize readout fidelity. The measurement pulse for another feedback cycle or final readout reaches the cavity at $t=2.8~\us$.}
\end{figure*}

\section{Measurement of the transmon populations}
\begin{figure}
\includegraphics[width=\columnwidth]{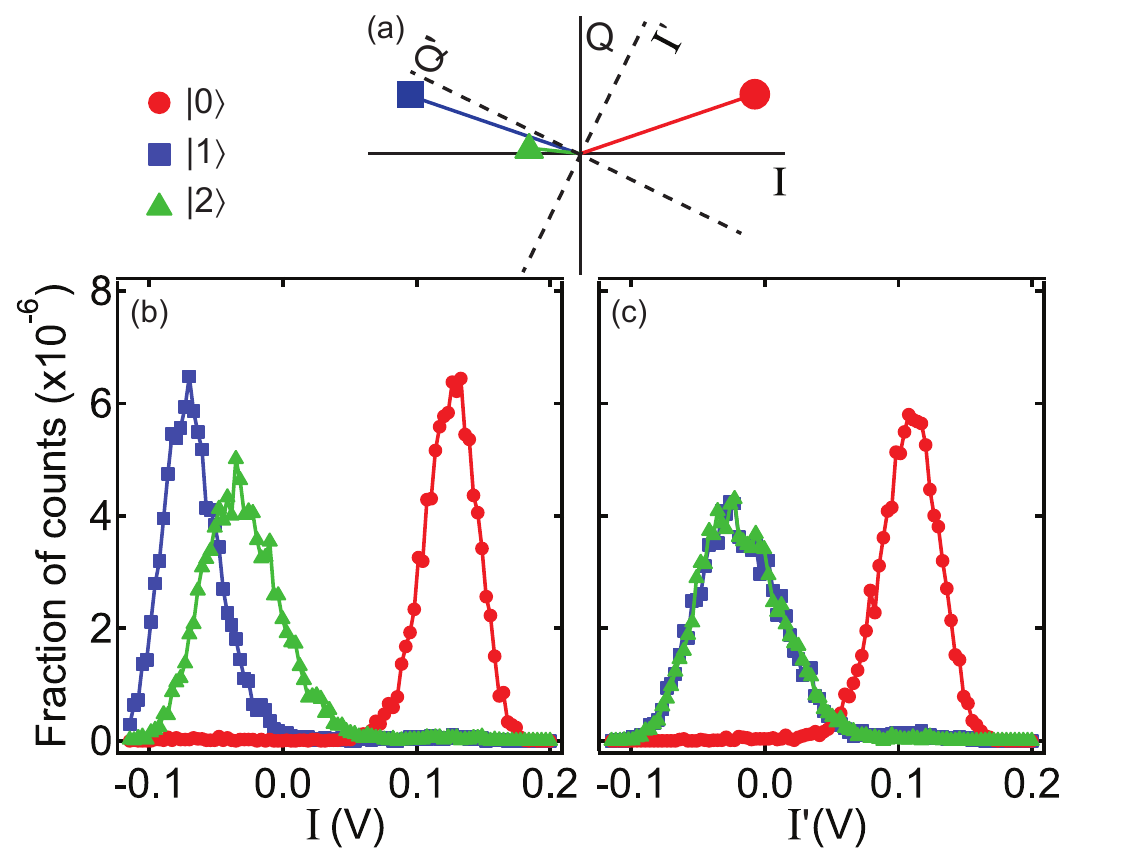}
\caption{The two readout configurations employed  and the associated histograms for the three lowest-energy transmon levels. (a) Sketch in phase space of the coherent states corresponding to the first three transmon levels. The amplified quadrature, chosen by tuning the RF-pump phase difference, is either I or $\mathrm{I'}$, giving the histograms in (b) and (c), respectively. (b) Optimum discrimination between $\ket{0}$ and $\ket{1}$. This choice is used in $\Ma$ and in all the measurements which are part of a feedback sequence to maximize the single-shot readout fidelity. (b) Readout configuration making $\ket{1}$ and $\ket{2}$ indistinguishable, as their respective coherent states have equal projection onto the $\mathrm{I'}$ axis. This configuration is used in the final measurement of every sequence to accurately extract the level populations, as described in the text.}
\end{figure}
Whereas readout for feedback maximally discriminates between $\ket{0}$ and $\ket{1}$ from single measurements, the final readout, used to quantify the reset error $\Perr$, relies on the average of thousands of runs. Due to leakage from the qubit subspace, the output voltage $\Vh$ also depends on $\Pee$. The average measurement is related to the populations as~\cite{Bianchetti10}
\begin{figure*}
\includegraphics[width=2\columnwidth]{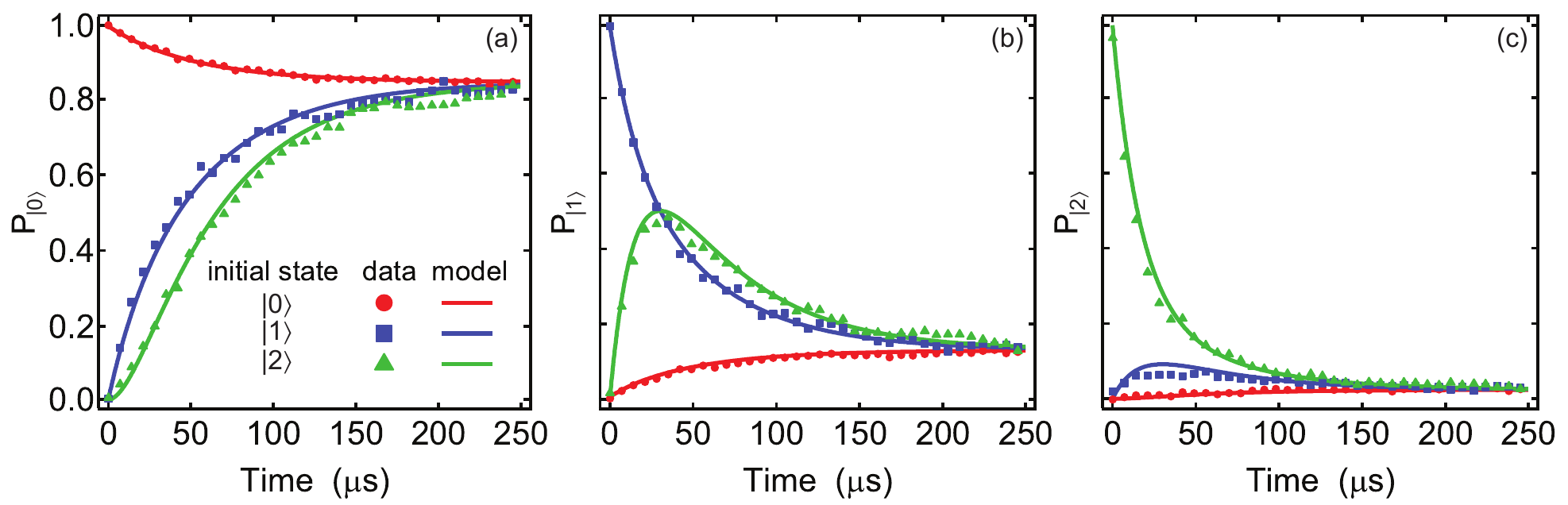}
\caption{Transmon population dynamics showing residual excitation. Populations (a) $\Pg$, (b) $\Pe$, and (c) $\Pee$ as a function of time following preparation in $\ket{0}$ (red), $\ket{1}$ (blue), and $\ket{2}$ (green). The combined fit of the master equation model to these data gives the transition rates and steady-state populations shown in Fig.~1.}
\end{figure*}
\begin{equation}
\label{eq:readout_2eq}
\avg{\Vh} =  \avg{\Vg} \Pg +  \avg{\Ve} \Pe +  \avg{\Vee} \Pee,
\end{equation}
 where the $\avg{V_{\ket{i}}}$ are the mean values of the distributions in Fig.~S3. 
When $\avg{\Ve} \neq \avg{\Vee}$, as in Fig.~S3(a), the measurement of $\avg{\Vh}$ is not sufficient to extract the three populations. Two possible strategies to obtain two additional equations are: 1) repeating the sequence, but swapping two of the populations prior to measurement 
in order to obtain different $\avg{V'_\mathrm{H}}$, $\avg{V''_\mathrm{H}}$; 2) analyzing the full time-domain response of the cavity~\cite{Bianchetti10}. We use a different approach, taking advantage of the degree of freedom in the relative phase between RF and pump, to extract $\Pg$ with the minimum number of measurements. As shown in Fig.~S3(b), the phase between RF and pump is chosen to maximize overlap between the histograms for $\ket{1}$ and $\ket{2}$. Making $\avg{\Ve}=\avg{\Vee}$, and assuming $\sum_{i=0}^2P_{\ket{i}}=1$ (as justified below), simplifies Eq.~(\ref{eq:readout_2eq}) to
\begin{equation}
\label{eq:readout_1eq}
\avg{\Vh} =  \Pg \avg{\Vg} + (1-\Pg) \avg{\Ve},
\end{equation}
with $\Pg$ as the only unknown. To measure $\Pe$, we transfer this population to $\ket{0}$ with a $\pi$ pulse, making $\avg{\Vh} = \Pe \avg{\Vg} + (1-\Pe) \avg{\Ve}$. Similarly,
to measure $\Pee$, we transfer this population to $\ket{0}$ with two consecutive $\pi$ pulses, the first on $\ottrans$ and the second on $\zotrans$. 
The amplitudes $\avg{\Vg}, \avg{\Ve}$ and $\avg{\Vee}$ are calibrated by preparing the corresponding states. To do this, we initialize to $\ket{0}$ as best we can ($\Pg=99.3\%$) by performing a measurement (used for postselection~\cite{Riste12,Johnson12}) and then applying the required $\pi$ pulses. Values for $\avg{\Vg}$, $\avg{\Ve}$ and $\avg{\Vee}$ are estimated from $\sim10\,000$ shots each. As shown in Fig.~S3(b), $\avg{\Ve}$ and $\avg{\Vee}$ are matched to better than $1\%$. The sum of the extracted populations shown in Figs.~S4(a-b) averages to $99.7\%$ and $99.5\%$ over the temporal evolutions starting from $\ket{0}$ and $\ket{1}$, respectively. For initial state $\ket{2}$ [Fig.~S4(c)], leakage to the third excited state is relevant under $50~\us$. In most cases, however, $\Pee$ is limited to a few percent, justifying the truncation of the model to the lowest three transmon levels.

\section{Model for measurement and feedback}
The readout errors $\epsilon^M_{i}$, indicating the probability of a wrong result $M$ for the pre-measurement state $\ket{i}$, are extracted from the histograms in Figs.~2 and S3, using the result of the fit in Fig.~S4 to account for the transmon equilibration between $\Ma$ and $\Mb$~\cite{Riste12}. The data yield the values $\epsilon_0^L=0.3\%, \epsilon_1^H=0.7\%$. By analogy with Ref.~\onlinecite{Riste12}, we approximate the errors with $\epsilon^L_0\sim p^L_{00}$ and $\epsilon^H_1\sim p^H_{01}$ (notation defined in the main text). Combining these with the best-fit values for Fig.~S3, $p^L_{00}+p^H_{10} = 0.4\%$ and $p^H_{11}+p^L_{01}=1.4\%$, we estimate the remaining errors $p^H_{10}=0.1\%$, $p^L_{01}=1.0\%$.

\begin{figure}
\includegraphics[width=\columnwidth]{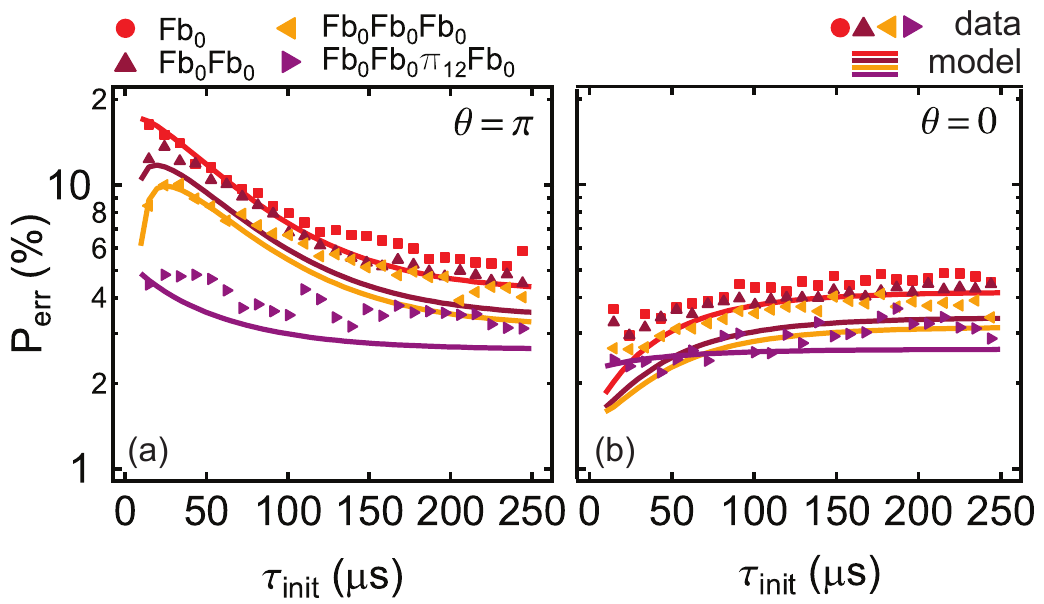}
\caption{Fast reset using multiple instances of $\Fbg$. The measurement sequence is illustrated in Fig.~4, with (a) $\theta = \pi$, and (b) $\theta=0$. At any $\tinit$ $\Perr$ decreases by a factor $\sim 3$ with one instance of $\Fbg$ compared to passive initialization (Fig.~4). Two or three iterations add only a modest improvement. An additional decrease in error by a factor $\sim 2$ at short $\tinit$ is obtained with a $\pi$ pulse on the $\ottrans$ transition preceding the third instance (purple). This pulse corrects for the leakage of population to $\ket{2}$, occurring for short $\tinit$. The last feedback round brings this population back to $\ket{0}$. Solid curves are given by the model, using the parameters extracted from Figs.~1 and 2 and assuming perfect pulses.}
\end{figure}

\section{Comparison of feedback protocols}
A double round of feedback tends to equalize the initialization error for different input states (Figs.~3 and S5). Augmenting the protocol to three or more rounds only marginally affects the residual difference, which is mainly due to leakage into $\ket{2}$. When a $\pi$ pulse is repeatedly applied on the $\zotrans$ transition [Figs.~3(a) and S5(a)], this leakage depends on the fraction of time that $\ket{1}$ is populated, which increases at shorter $\tinit$. Therefore, $\Pee$ can exceed its steady-state value, leading to increased $\Perr$. We address this problem by inverting the populations in $\ket{1}$ and $\ket{2}$ after two iterations of $\Fbg$. This brings the population, previously in $\ket{2}$, within reach of a third instance of $\Fbg$. This strategy is only effective when $\Pee>\Pe$, explaining why a residual error of $\sim3\%$ is observed at long $\tinit$, when, after two rounds of feedback, $\Pe\approx\Pee$. Similarly, the repeated initialization in $\ket{0}$ without periodically applying a $\pi$ pulse [Figs.~4(b) and S5(b)] limits the escape probability to the excited states, as evident from the reduced error at short $\tinit$.